\documentstyle[prl,aps,multicol,epsf]{revtex}
\def \be{\begin{equation}}
\def \ee{\end{equation}}
\def \k{{\bf k}}
\def \om{\omega}
\begin{document}

\bibliographystyle{simpl1}

\title{Comment on ``Temperature Dependence of the Josephson Current''}

\author{Vinay Ambegaokar}
\address{Laboratory of Atomic and Solid State Physics,
Cornell University, Ithaca NY 14853, USA}
\maketitle
\centerline{\today}
\bigskip
In a recent letter \cite{OV}, Overhauser has claimed that the accepted
calculations of the temperature dependence of the Josephson
current \cite{BDJ,AB} are incorrect.  His contention is that broken pair
states, in the BCS \cite{BCS} terminology, should not be included as
initial
states in calculating the contribution of tunneling to the free energy of
two weakly coupled superconductors.\cite{ETC}

In fact, all terms that depend on the difference of the phases of the order
parameters in the two superconductors must be included and such phase
dependences are present (via intermediate states) in the omitted terms. As
Overhauser himself shows, including them reproduces the usual result.

This is really all that needs to be said.  The following remarks may,
however, clarify the issue---if further clarification is needed.

Consider the anomalous averages
\be
\label{1}
F^>(\k,t,t^{\prime})\equiv \langle a_{\k\uparrow}(t)
a_{-\k\downarrow}(t^\prime)
\rangle,~{\rm and}~F^<(\k,t,t^{\prime})\equiv \langle
a_{-\k\downarrow}(t^\prime)a_{\k\uparrow}(t) \rangle.
\ee
The difference between the conventional ($C$) and Overhauser ($O$) way of
evaluating these averages may be summarized in the following relations for
their Fourier transforms with respect to $t - t^\prime$:
\be
\label{2}
F_C^>(\k,\om )=B(\k,\om )[1-f(\om )],~~~~~~~F_C^<(\k, \om)=B(\k,\om )f(\om
),
\ee
as opposed to
\be
\label{3}
F_O^>(\k,\om )=B(\k,\om )[1-f(\om )]^2,~~~~~~F_O^<(\k, \om)=B(\k,\om )f(\om
)^2.
\ee
Above, $B(\k,\omega)$ is an odd function of $\om$ whose form is not
essential for the argument and $f(\om )=[\exp(\beta\om ) + 1]^{-1}$ is
the Fermi function, with $\beta$ the reciprocal of the temperature.

Note that the conventional calculation yields forms which agree with the
general fluctuation-dissipation requirement
\be
\label{4}
F^>(\om )= e^{\beta \om} F^<(\om ),
\ee
whereas these quantitites calculated according to the Overhauser
prescription do not, showing that the ad hoc suppression of states
(which treats initial and intermediate states asymmetrically) is
unphysical.

Work supported in part by the NSF under grant DMR-9805613.


\begin{references}
\bibitem{ETC} Other assertions in ref. \cite{OV} of error are also
without merit.  The use of the Heisenberg representation is an
unobjectionable matter of
taste.  The `factorization' is not an
approximation:  to second order in the tunneling, the average is in a
product of uncoupled left and right thermal ensembles.
\end{references}
\end{document}